\documentclass[fleqn, 10pt]{wlscirep}
\usepackage{graphicx}
\usepackage{amssymb}
\usepackage{amsmath}
\usepackage{multirow}
\usepackage{gensymb} 
\usepackage{float}
\title{Anomalous anisotropic behaviour of spin-triplet proximity effect in Au/SrRuO$_3$/Sr$_2$RuO$_4$ junctions}
	 
	\author[1,2,*]{M.~S.~Anwar}
	\author[1]{M.~Kunieda}
	\author[3,4]{R.~Ishiguro}
	\author[5,6]{S.~R.~Lee}
	\author[1]{C.~Sow}
	\author[2] {J.~W.~A.~Robinson}
	\author[1]{S.~Yonezawa}
	\author[5,6]{T.~W.~Noh}
	\author[1]{Y.~Maeno}
\affil[*]{Corresponding author: msa60@cam.ac.uk}  
\affil[1]{Department of Physics, Kyoto University, Kyoto 606-8502, Japan} 
\affil[2]{Department of Materials Science and Metallurgy, Cambridge University, Cambridge U.K.}
 \affil[3]{Department of Mathematical and Physical Sciences, Japan Women's University, Tokyo 112-8681, Japan} 
 \affil[4]{Department of Applied Physics, Tokyo University of Science, RIKEN Tokyo 162-8601, Japan}
\affil[5]{Center for Correlated Electron Systems, Institute for Basic Science (IBS), Seoul 151-747, Republic of Korea}
\affil[6]{Department of Physics and Astronomy, Seoul National University, Seoul 151-747, Republic of Korea}

\begin{abstract}

Spin-polarized supercurrents can be generated with magnetic inhomogeneity at a ferromagnet/spin-singlet-superconductor interface. In such systems, complex magnetic inhomogeneity makes it difficult to functionalise the spin-polarized supercurrents. However, spin-polarized supercurrents in ferromagnet/spin-triplet-superconductor junctions can be controlled by angle between magnetization and spin of Copper pairs ($d$-vector), that can effectively be utilized in developing of a field of research known as superconducting spintronics. Recently, we found induction of spin-triplet correlation into a ferromagnet SrRuO$_3$ epitaxially deposited on a spin-triplet superconductor Sr$_2$RuO$_4$, without any electronic spin-flip scattering. Here, we present systematic magnetic field dependence of the proximity effect in Au/SrRuO$_3$/ Sr$_2$RuO$_4$ junctions. It is found that induced triplet correlations exhibit strong anisotropic field response. Such behaviour attributes to the rotation of the $d$-vector of Sr$_2$RuO$_4$. This anisotropic behaviour is in contrast with the vortex dynamic. Our results will stimulate study of interaction between ferromagnetism and unconventional superconductivity.

\end{abstract}

\begin{document}

\flushbottom
\maketitle

\section{Introduction}

Generation of dissipationless spin-polarized (spin-triplet) supercurrent is the major interest of superconducting devices, which can be utilized to establish energy efficient superconducting spintronics~\cite{Eschrig2015,Linder2015}. In last two decades, rigorous works have been conducted to produce and control spin-triplet supercurrents using heterostructures of ferromagnets (Fs) and conventional spin-singlet superconductors (SSCs)~\cite{Bergeret2001,Keizer2006,Eschrig2008,Khaire2010,Robinson2010,Anwar2010,Anwar2012,Bergeret2013,Bergeret2014,Jacobsen2016,Banerjee2017}.  It has been established that magnetic inhomogeneity is always required to emerge spin-triplet supercurrent at F/SSC interface, which can be achieved by using non-collinear magnetization in multilayer ferromagnets~\cite{Khaire2010,Robinson2010,Anwar2012}. Complicated magnetic structure of multilayer ferromagnets makes it hard to functionalize the F/SSC devices. It can be settled by replacing SSC with a spin-triplet superconductor (TSC). In F/TSC heterostructures, a single F layer can affectively emerge spin-polarized supercurrents with fully conserved spin degree of freedom in the entire device. Furthermore, recent theoretical work suggested that superconducting properties of F/TSC junctions strongly depends on relative orientation of magnetization ($m$) of a F and $d$-vector of a TSC~\cite{Brydon2009a,Brydon2009b,Annunziata2011,Gentile2013,Terrade2013}. Interestingly, when the $d$-vector and $m$ are perpnadicular (angle between $d$-vector and $m$ $\theta_{md}=\pi/2$)  the spin-triplet correlation can be induced monotonically over a long range. However, for parallel configuration, $\theta_{md}=0$, the induced order parameter oscillates spatially~\cite{Terrade2013}.

Sr$_2$RuO$_4$ (SRO214) is one of the best-candidates of TSCs~\cite{Maeno2012} with the superconducting critical temperature ($T_{\rm c\text{-}bulk}$) of 1.5~K. Most likely, it exhibits the chiral $p$-wave spin-triplet state with broken time-reversal symmetry~\cite{Maeno1994,Luke1998,Ishida1998,Nelson2004,Xia2006,Kindwingira2006,Nakamura2011,Anwar2013,Anwar2017}, although there are still unresolved issues~\cite{Hicks2010,Yonezawa2013,Hassinger2017}. Very recent nuclear magnetic resonance (NMR) study shows the reduction of Knight shift, which cannot be explained by spin-striplet scenario\cite{Pustogow2019,Ishida2019}. Recently, SRO214 attracts interest for exploring topological superconducting phenomena originating from its orbital phase winding~\cite{Nakamura2011,Maeno2012,Anwar2013,Anwar2017}. The superconducting order parameter of a SRO214 (TSC) can be represented by a vector ${d}=\hat{\bf z} (p_x \pm ip_y)$, where $\hat{\bf z}$ is the out-of-plane (along the $c$-axis) basis vector, and $P_x$ and $P_y$ are the $x$ and $y$ components of the orbital order parameters, respectively~\cite{Maeno2012}. Note that the $d$-vector is always perpendicular to the spin of triplet Cooper pairs. For bulk SRO214 superconductor $d$-vector is aligned along the $c$-axis (out-of-plane) fixed by spin-orbit coupling~\cite{Ng2000a,Ng2000b,Annett2006} but at the surface it may not be true because of surface effects. It is expected that $d$-vector may rotate to the in-plane direction with external magnetic field of about 20~mT applied along the out-of-plane direction~\cite{Murakawa2004,Annett2008}.  

Newly, we developed F/TSC heterostructures by growing epitaxial ferromagnetic SrRuO$_3$ (SRO113) thin films on a spin-triplet superconductor SRO214 single crystal using~pulsed laser deposition~\cite{Anwar2015}. Furthermore, long-ranged proximity effect is also observed, where even frequency $p$-wave spin-triplet may dominate compared to odd frequency $s$-wave spin-triplet, as junctions were in clean limit~\cite{Anwar2016, Anwar2019}.    

In this article, we present our investigations on differential conductance ($dI$/$dV$) of Au/SRO113/SRO214 junctions as a function of temperature and magnetic field applied along both in-plane ($H_{\rm in}$) and out-of-plane ($H_{\rm out}$) directions. We found that the induced correlation exhibits anisotropic behavior in the response of applied field. Vortex dynamics cannot explain our results. This anisotropic effect can be attributed to the relative orientation of the $d$-vector of SRO214 and $m$ of SRO113.

\begin{figure}
	\begin{center}
		\includegraphics[width=8cm]{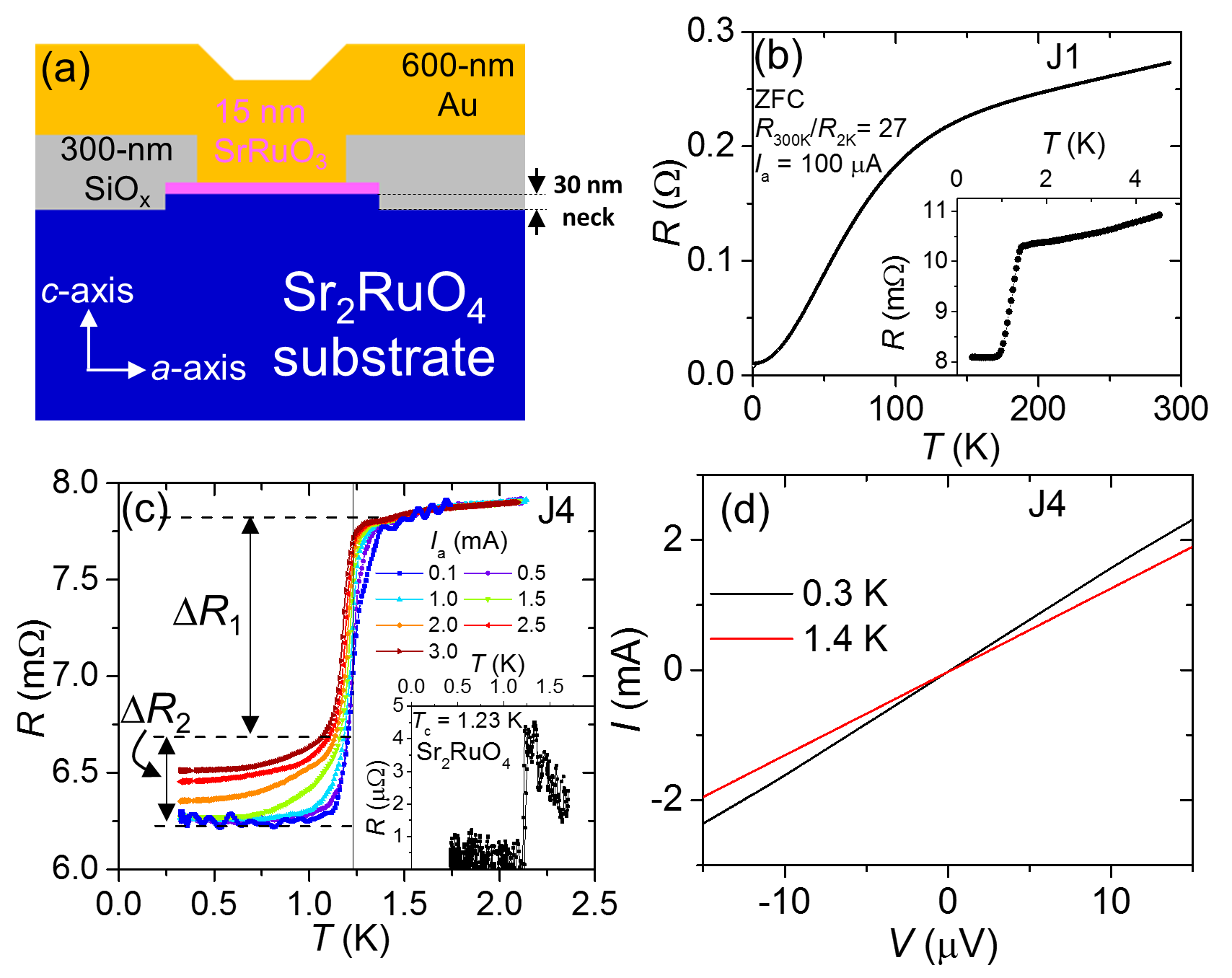}
		\caption{(a) Schematic of Au/SrRuO$_3$/Sr$_2$RuO$_4$ junctions. Note that a $\approx$30-nm thick neck of SRO214 was prepared in order to separate a part of superconductor from the bulk substrate. (b) Temperature dependent resistance $R$($T$) from 300~K down to 2~K for a Junction J1 with the junction area of 20 $\times$ 20 $\mu$m$^2$. Inset shows $R$($T$) close to the superconducting transition. (c) $R$($T$) of Junction J4 (5 $\times$ 5 $\mu$m$^{2}$) at various applied currents. Vertical solid line indicates the bulk $T_{\rm c}$ of SRO214. Inset shows $R$($T$) of SRO214 substrate measured by using a four probes technique. (d) Current-voltage curves below and above $T_{\rm c}$ of Junction J4.}
		\label{device_RT}
	\end{center}
\end{figure}

\begin{figure*}
	\begin{center}
		\includegraphics[width=16cm]{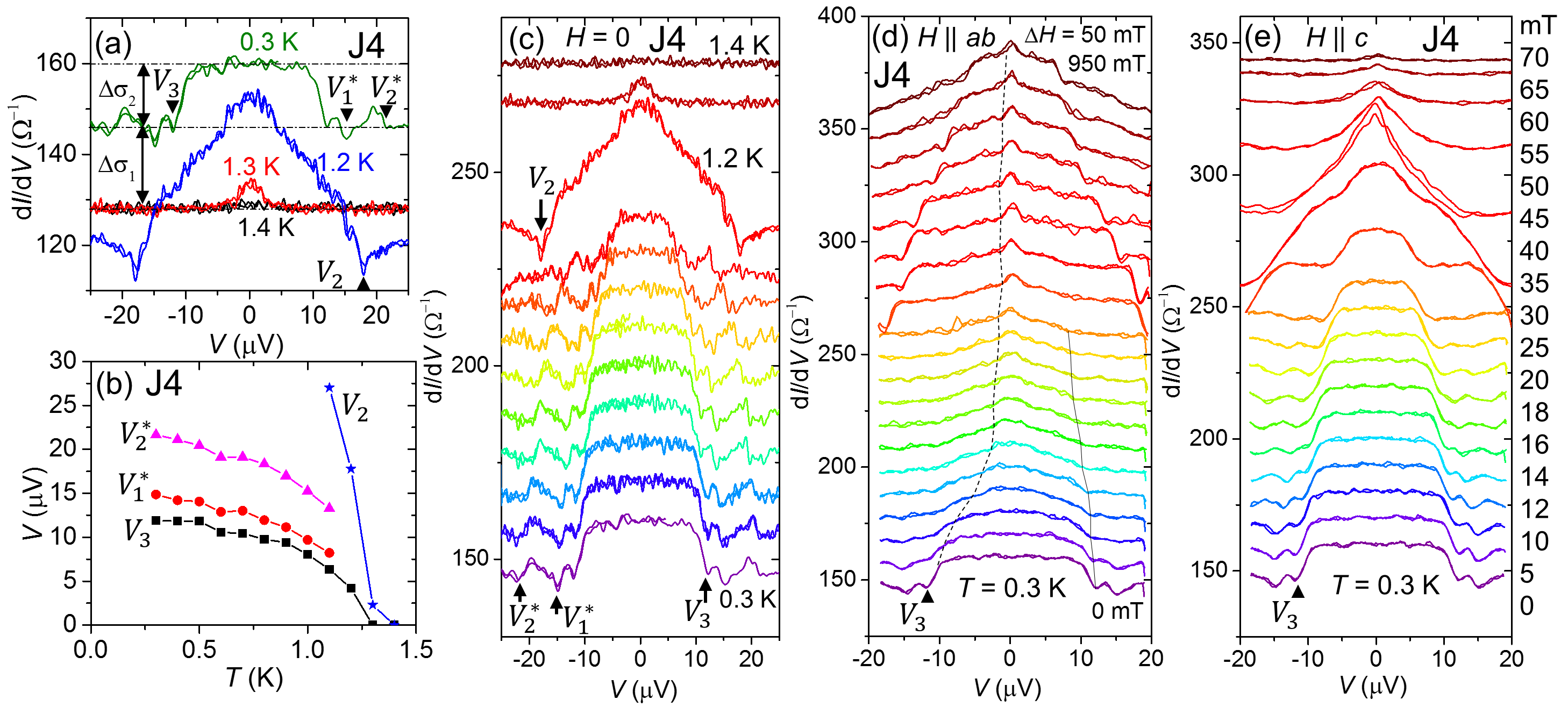}
		\caption{Temperature and field dependent differential conductance ($dI$/$dV$) of Au/SrRuO$_3$/Sr$_2$RuO$_4$ Junction J4. (a) $dI$/$dV$ measured at different temperatures between 1.4~K and 0.3~K. At 0.3~K a flat-top central conductance peak appears between $\pm V_3$. Note that there are additional oscillations with the minima at $V_1^*$ and $V_2^*$. At temperatures above 1.1~K, another transition at $\pm V_2$ becomes visible within the range of bias voltage. (b) Temperature dependence of all characteristic voltages appeared in $dI$/$dV$ data. (c) $dI$/$dV$ at various temperatures. The data are shifted with a step of 10 m$\Omega^{-1}$ but for 1.2~K and 1.3~K are shifted with 30 m$\Omega^{-1}$ for clarity. $dI$/$dV$ as a function of magnetic field applied along the in-plane (d) and the out-of-plane ($c$-axis) directions. Data are shifted for clarity.}
		\label{IVT_J4}
	\end{center}
\end{figure*}

\section{Results}

In this article, we present the results of various junctions but mainly focused on two junctions J4 and J5 that exhibit different normal-state resistance ($R_{\rm N}$) 7.83 m$\Omega$ and 198 m$\Omega$, respectively. A schematic illustration of a junction is shown in the Fig. \ref{device_RT}(a). 

Temperature dependent resistance $R$($T$) was measured during cooling in zero field down to 300~mK. SRO214, SRO113 and Au are good metals with very low resistivity at low temperatures, except the resistivity of SRO214 along the $c$-axis ($\rho_c$)~\cite{Hicks2010}. It suggests that at low temperatures resistivity of our junctions is dominated by the SRO113/SRO214 and Au/SRO113 interfaces~\cite{Anwar2015,Anwar2016}. Our present junctions exhibit RRR = 27 that is three times larger than that of our previous junctions~\cite{Anwar2016}. Furthermore, a sharp superconducting transition is observed that reflects good quality of the junctions (inset of Fig.~\ref{device_RT}(b)). It is also clear that in the normal state, the resistivity data have major~contributions of $\rho_c$ of SRO214. It indicates that current is flowing along the normal of the junctions, which excludes the possibility of any direct contact between Au electrode and SRO214~superconductor.

Figure~\ref{device_RT}(c) presents $R$($T$) at low temperatures of Junction J4 (5 $\times$ 5 $\mu$m$^2$) measured with different applied currents ($I_{\rm a}$ in the range of 0.1 to 3 mA) that shows two superconducting regions robust and weak against $I_{\rm a}$. At $I_{\rm a}$~=~0.1~mA the transition onset appears at $T_{\rm c}~=$~1.38~K, which is higher than the $T_{\rm c}~\approx$~1.23~K of SRO214 substrate (see the inset of Fig.~\ref{device_RT}(c)). This increase in $T_{\rm c}$ may arises due to mutual strain (pressure) between the substrate and the film close to the interface~\cite{Hicks2017}. The $R$($T$) behavior with increasing $I_{\rm a}$ reveals that there are two distinct transition regions. The first transition with change in the resistance $\Delta R_1~=$~1.15~m$\Omega$ is only weakly dependent on $I_{\rm a}$. The second transition is strongly suppressed with increasing $I_{\rm a}$ with the change in the resistance of $\Delta R_2=$ 0.45~m$\Omega$. Multiple transitions are expected to emerge in a multi-barrier junction such as the junctions presented here: Normal-metal/F/TSC. Essentially, the same behaviour has been observed in our previous study~\cite{Anwar2016}. These observations suggest that $\Delta R_1$ and $\Delta R_2$ correspond to SRO113/SRO214 and Au/SRO113 interfaces, respectively. It means that the $dI$/$dV$ as a function of bias voltage should exhibit Andreev reflection (AR) features occurring at two distinct interfaces. Note that AR can occur for a bias voltage below superconducting gap (induced minigap in SRO113 \cite{Anwar2019}) at SRO113/SRO214 (Au/SRO113) interface. To study the AR, we obtained $dI$/$dV$($V$) by taking first order numerical derivative of the measured current-voltage ($I$-$V$) curves shown in Fig.~\ref{device_RT}(d). We investigate the $dI$/$dV$ as a function of externally applied magnetic fields both in-plane $H_{\rm in}$ and $H_{\rm out}$ directions and temperature as well.

\begin{figure*}
	\begin{center}
		\includegraphics[width=14cm]{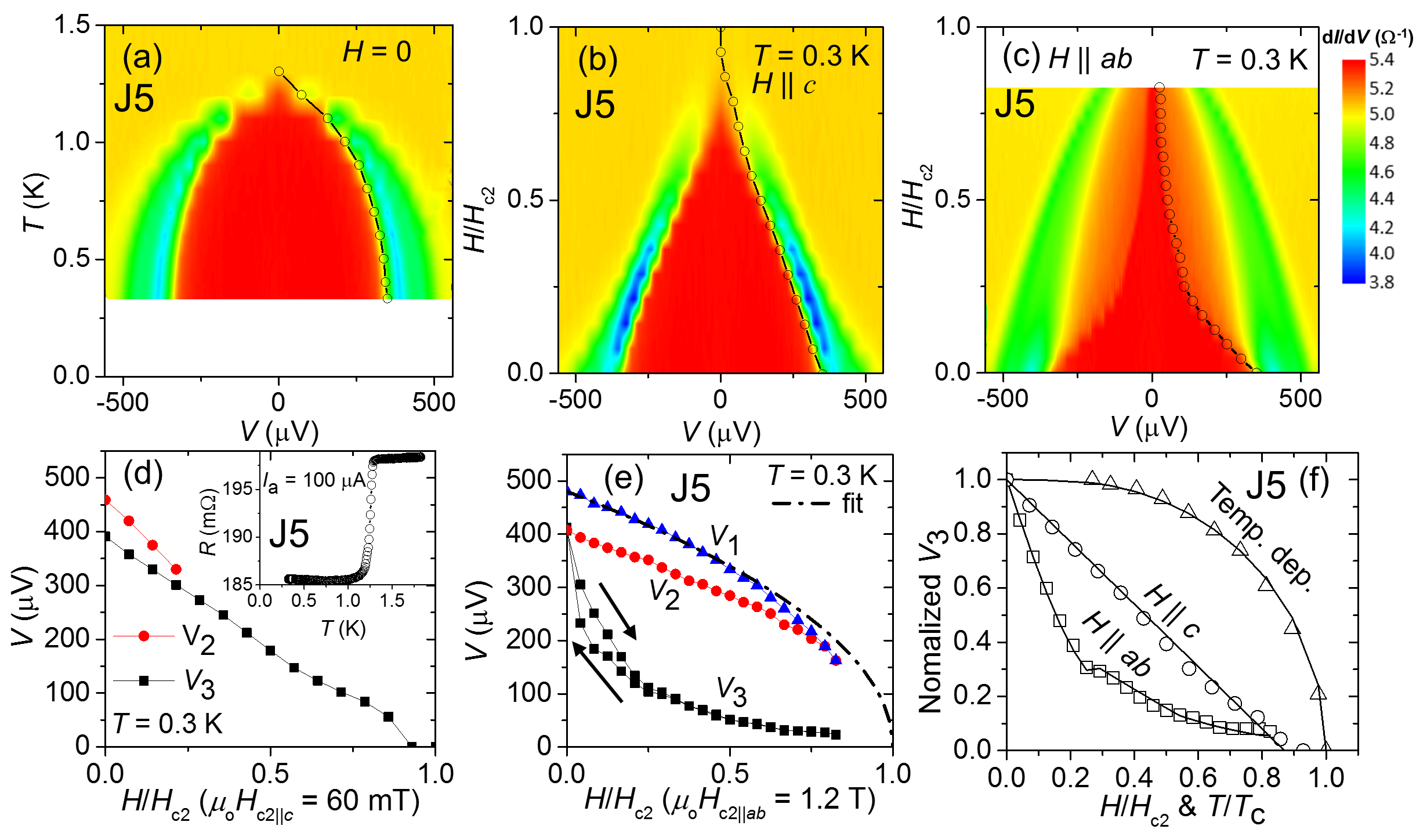}
		\caption{Differential conductance of Au/SrRuO$_3$/Sr$_2$RuO$_4$ Junction J5. (a) Colour plot of $dI$/$dV$($V$) vs temperature. The $V_3$ is indicated with open circles. $dI$/$dV$($V$) vs magnetic field applied along (b) the $c$-axis and (c) the in-plane directions, respectively. (d) $V_2$ and $V_3$ vs applied field along the $c$-axis normalized by the upper critical field. Inset shows $R$($T$) at low temperatures of Junction J5. (e) All characteristic voltages $V_1$, $V_2$ and $V_3$ vs in-plane applied field. $V_1$ and $V_2$ decrease expectedly. In contrast $V_3$ exhibits an anisotropic behavior with hysteresis below 200~mT. (f) Normalized $V_3$ vs applied field along both in-plane ($H_{\rm in}$) and out-of-plane directions $H_{\rm out}$, as well as vs temperature plotted for comparison.}
		\label{IVTH_J5}
	\end{center}
\end{figure*}

First, we present $dI$/$dV$ data measured at different temperatures (see Fig.~\ref{IVT_J4}(a)). It shows an enhancement in $dI$/$dV$ below superconducting transition temperature $T_{\rm c}$. Below $T_{\rm c}$, the conductance of a transparent F/SSC interface increases with opening of AR channel, which is related with spin-polarization of the ferromagnet as well. At 0.3~K and zero applied field (after zero field cooling), the conductance enhances by $\approx$~10$\%$ below $\pm V_3$ with almost flat top. This transition yields the change in the conductance $\Delta\sigma_2~=$~13.84~$\Omega^{-1}$ that corresponds to $\Delta R_2$ (second transition in $R$($T$)). At temperatures higher than 1.1~K an additional enhancement within our applied bias voltage can be observed at $\pm V_2$. Note that there are sharp dips associated with $V_2$ features (Fig.~\ref{IVT_J4}(a)), which may correspond to the destruction of superconductivity at SRO113/SRO214 interface with critical current\cite{Anwar2019}. However, the temperature dependence of $dI$/$dV$ (Fig.~\ref{IVT_J4}(c)) shows that the dip persists up to the bulk $T_{\rm c}$, which suggests that the dip can be attributed to the critical current of SRO214-neck part of the junctions, and the proximity effect at the SRO113/SRO214 interface is simultaneously switched off. The $V_2$ transition exhibits $\approx$~14$\%$ increase with $\Delta\sigma_1~=$~18.04 $\Omega^{-1}$ consistent with $\Delta R_1$ (first transition in $R$($T$)). It indicates that characteristic voltages of $V_3$ and $V_2$ are corresponding to multi AR features occurred at Au/SRO113 and SRO113/SRO214 interfaces, respectively. The data presented in Fig.~\ref{IVT_J4}(b)) is extracted from Fig.~\ref{IVT_J4}(c). It shows a monotonic suppression of the AR with increase in the temperature. Similar behaviour is observed for junction J5 as well which is shown in Fig.~\ref{IVTH_J5}(f) and discussed more at the end of the section.

The observations of multiple enhancements in $dI$/$dV$ reveal that superconductivity penetrates into a 15-nm thick SRO113 layer and reaches at Au/SRO113 interface as well, which is only possible with spin-triplet superconducting correlation, since spin-singlet coherence length of SRO113 is on the order of 1-nm~\cite{Asulin2006}. Note that two additional oscillations with minima at $V_1^*$ and $V_2^*$ appear between $V_3$ and $V_2$. These oscillations are suppressed monotonically just like $V_3$ with the increase in the temperature, see Fig.~\ref{IVT_J4}(b) and (c). These oscillations may correspond to McMillan-Rowell resonance~\cite{Visani2012} resulting from interference between Andreev reflected quasiparticles at the SRO113/SRO214 interface and the quasiparticles reflected back into SRO113-layer from the Au/SRO113 interface.

The most interesting and important part of this study is voltage biased $dI$/$dV$ as a function of applied field. Figures~\ref{IVT_J4}(d) and (e) shows the $dI$/$dV$($V$) data obtained with the fields $H_{\rm in}$ and $H_{\rm out}$, respectively. In case of the applied field $H_{\rm in}$, $V_3$ is abruptly suppressed up to 200~mT and decreases slowly with further increase in $H_{\rm in}$. In contrast $V_2$ is suppressed gradually, which is an expected behaviour of Cooper pair breaking under applied magnetic fields (orbital effects). For fields $H_{\rm out}$, both $V_3$ and $V_2$ decreases monotonically. Thus spin-triplet proximity effect into the SRO113 layer exhibits an anisotropic response to applied field. Such an effect can be expected for anisotropic $d$-vector rotation with applied field, further discussions are given in the discussions section.

Junction J4 with lower $R_{\rm N}$ exhibits various interesting features in the junction voltage. However, concerning to the signal to noise ratio a junction with larger $R_{\rm N}$ is preferable. Indeed in Junction J5 with rather high $R_{\rm N} = $ 198.2~m$\Omega$, we observed good quality data that also exhibits AR features with multiple transitions. For Junction J5, $dI$/$dV$($V$) as a function of temperature and applied field exhibit two transitions corresponding to $V_3$ and $V_2$. Note that we also obtained the $V_1$ feature particularly at 0.3~K and zero field as observed in our previous junctions~\cite{Anwar2016}. The characteristic voltage $V_1$ most probably arises due to critical-current of SRO214 neck~\cite{Anwar2016}. Note; in this junction the oscillations corresponding to McMillan-Rowell resonance are not observed due to its lower amplitude and higher $R_{\rm N}$. By comparing Figs.~\ref{IVTH_J5}(b) and (c), a strong anisotropy in $V_3$ is obviously reproduced. In fact, we observed this anisotropic effect in almost all our working junctions. 

For Junction J5, under the effect of $H_{\rm out}$ fields, the characteristic voltage $V_2$ shown in blue Fig.~\ref{IVTH_J5}(b) becomes vague at higher fields as shown in Fig.~\ref{IVTH_J5}(d). It reflects that in junctions with higher $R_N$ (lower interface conductance) the AR features are overlapped with each other. However, all the characteristic-voltage features are obviously present for all $H_{\rm in}$ fields, see Fig.~\ref{IVTH_J5}(e). Interestingly, $V_3$ exhibits hysteretic effect only below 200 mT. For a comparison, we plotted normalized $V_3$ as a function of field and temperature in Fig.~\ref{IVTH_J5}(f). 

We applied theoretical fits on temperature and field dependent $V_3$ to analyse its behavior. The behaviour of $V_3$ versus temperature follows the interpolation formula, $V_3(T) = V_3(0){\rm tanh}\sqrt{a(T_c/T-1)}$ with constant $a=1.56$, which is relatively lower than the expected value of 1.74 for s-wave conventional BCS superconductivity. Interesting, a=1.56 is indicating
unconventional superconducting proximity effect and potentially consistent with $p$-wave superconductivity \cite{Nomura2001,Anwar2019}. The monotonic suppression of $V_3$ under fields $H_{\rm out}$ can be reasonably fitted with the expression $V_3(H)=V_3(0)\sqrt{(1-H/H_c)}$. However, $V_3$ under fields $H_{\rm in}$ is strongly suppressed at lower fields less than 200~mT and slowly decreases for higher fields. Note that the coercive field of SRO113 layer is $\approx$ 200~mT for rectangular pads of the size of tens of microns~\cite{Anwar2016}. It indicates that there are two distinct decoherence effects on the induced spin-triplet correlation in the SRO113 layer, such as orbital effect and $m$ rotation relative to the $d$-vector of SRO214 superconductor. The phenomenological fit shown in Fig. ~\ref{IVTH_J5}(f) is discussed below.  

\section{Discussion} 

Before starting our discussion, let us summarize our main results. We observed AR features in $dI$/$dV$ with characteristic voltages $V_3$ and $V_2$ emerged at Au/SRO113 and SRO113/SRO214 interfaces, respectively. The $V_3$ ($V_2$) exhibits anisotropic (isotropic) behaviour in the response of applied magnetic field. The AR feature with $V_3$ is suppressed strongly below 200~mT and decreases slowly with farther increase under the field $H_{\rm in}$. However, $V_2$ decreases monotonically with $H_{\rm out}$ and $H_{\rm in}$ as well. For the field along the $c$-axis $V_3$ decreases similar to $V_2$. 

\begin{figure}
	\begin{center}
		\includegraphics[width=16cm]{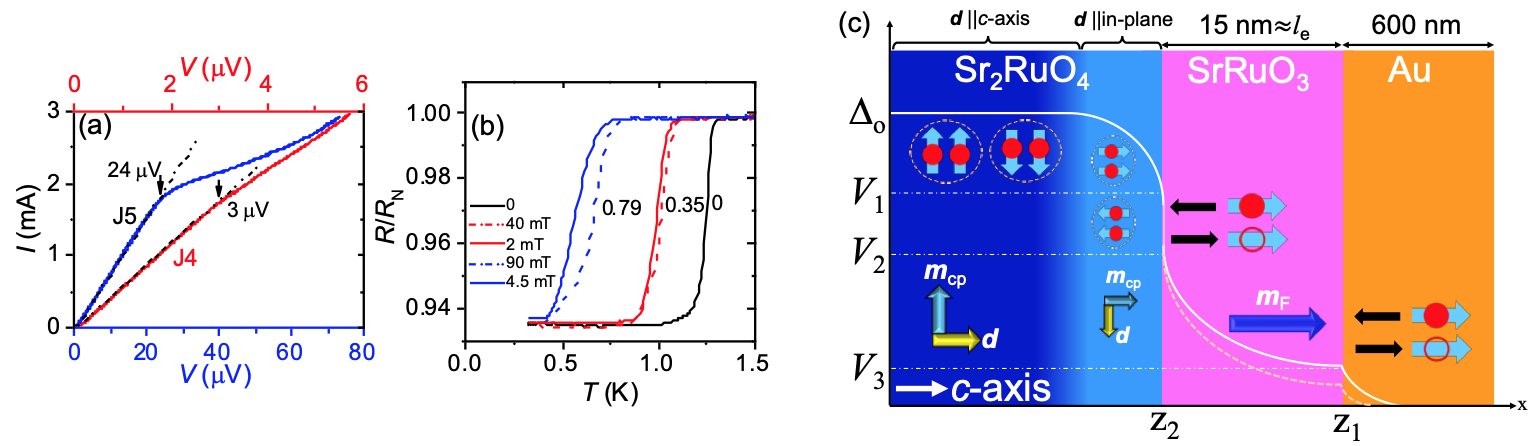}
		\caption{(a) Current-voltage curves at 0.3~K for Junctions J4 (red) and J5 (blue) after subtracting additional contributions. (b) $R$($T$) at different applied fields both $H_{\rm in}$ and $H_{\rm out}$. (c) Schematic illustration of induction of superconductivity in the junction. Explanations are given in the main text.}
		\label{IV_RT}
	\end{center}
\end{figure}

First, we discuss the subtraction of resistance contributions other than interfaces to estimate the accurate values of characteristic voltages $V_3$ and $V_2$. As described below, the measured resistance of our devices may contain some contributions from non-interface resistance, such as from the neck part of SRO214. To estimate accurate values of characteristic voltages specifically, $V_3$ at 0.3 K, we subtract the contributions of estimated additional resistances. For simplicity, let us assume an ideal situation that AR probability is 100$\%$ so that the conductance of each interface is enhanced by a factor of two compared to its normal-state conductance. Under this assumption, for Junction-J4 (Junction-J5) the total resistance of Au/SRO113 and SRO113/SRO214 interface is two times the $\Delta R = \Delta R_1 + \Delta R_2 = 1.55$~m$\Omega$ (12.74~m$\Omega$). At 0.3 K, additional resistance of Junction-J4 (Junction-J5) $R_a = R_N - 2\Delta R = $ 4.73 m$\Omega$ (172.72~m$\Omega$) that contributes additional voltage drop across the junction. We subtract a linear contribution of $R_a$ and plotted the results in Fig.~\ref{IV_RT}(a). It reveals that $V_3$ for junction-J4 (junction-J5) is $\approx$~3~$\mu$V ($\approx$~20~$\mu$V). It shows that junction-J5 exhibits an order higher $V_3$ due to lower interface transparency and more drop of voltage across it.

It is important to discuss the vortex effect on proximity effect in Au/SRO113/SRO214 junctions, which can also be anisotropic. Of course, vortices can be induced for the fields higher than $H_{c1}$. Note that $H_{c1}$ of SRO214 is 1~mT and 70~mT along the $a$-axis (in-plane) and the $c$-axis\cite{Maeno2012}, respectively. Transport properties of a superconducting junction can be affected when vortices are crossing the interface because of vortex dynamics. According to the geometry of our junctions ($c$-axis is out-of-plane normal of the junction), the field induced vortices can encounter the interface only for fields $H_{\rm out}$. However, vortices align parallel to the interface for the fields $H_{\rm in}$. It means, AR features should exhibit additional suppression for fields $H_{\rm out}$. In contrast, $V_3$ is suppressed strongly for the fields $H_{\rm in}$ only. 

To understand the effect of vortices in our junctions, we measured $R$($T$) under the fields $H_{\rm in}$ and $H_{\rm out}$. The $T_c$ of first transition with $\Delta R_1$ of SRO113/SRO214 interface decreases more for same effective field applied along the $c$-axis compared for the in-plane direction, see Fig.~\ref{IV_RT}(b). However, second transition corresponding to Au/SRO113 interface is in contrast. It indicates that the electronic transport properties of our junctions can encounter vortices only for the fields $H_{\rm out}$. Furthermore, it also reveals that the observed anisotropic response of $V_3$ under applied fields does not arise due to vortex dynamics. 

Effect from the spin part: spin-triplet proximity effect at F/TSC interface can be dictated by angle $\theta_{dm}$ between $d$-vector of SRO214 and $m$ vector of SRO113 layer~\cite{Brydon2009a,Brydon2009b,Annunziata2011,Gentile2013,Terrade2013}. We applied a following phenomenological fit to $V_3$($H_{ab}$),
\begin{equation}
V_3(H) = V_3(0)[F(H) - G(\theta_{md})]
\end{equation}
Where, $V_3$(0) is the characteristic voltage at zero temperature, $F(H) = \sqrt{1 - H/H_c}$ and $G(\theta_{md}) = A$cos($\theta_{md}$) with a constant $A$. We obtained a good fit by changing the $\theta_{md}$ nonmonotonically; first decreased until 200~mT and increased for higher fields. Furthermore, we presumed that $d$-vector is aligned along the in-plane direction at the interface.   

In bulk SRO214 superconductor, the $d$-vector is aligned along the $c$-axis and in the result the spins of the Cooper pairs are aligned along the $ab$-plane. Thus, the $d$-vector and Cooper pair spins can be rotated with an applied magnetic field $H_{\rm out}$ of 20~mT \cite{Murakawa2004,Annett2008} to in-plane and to out-of-plane, respectively. In this scenario, $H_{\rm in}$ cannot rotate the $d$-vector. That means, proximity effect should supress more in case of $H_{\rm out}$ rather than $H_{\rm in}$. But, our results are suggesting in contrast. 

If the $d$-vector is perpendicularto the $c$-axis particularly at the interface, $V_3$ can be affected with rotation of $\theta_{md}$ with $H_{\rm in}$. It can be possible that broken inversion symmetry at the SRO113/SRO214 interface induces Rashba-type~\cite{Rashba1960} spin-orbit coupling with in-plane vector characterizing the spin–orbit coupling $g=(k_x, k_y, 0)$ with out-of-plane effective field, which tends to align the $d$-vector along the in-plane direction. Additionally, out-of-plane magnetic anisotropy of SRO113 thin film is also supporting such an alignment of the $d$-vector. In the result, at zero applied field, spin of triplet Cooper pairs in SRO214 and $m$ of SRO113 are parallel and in the result $\theta_{md} = \pi/2$, which is a favourable configuration for proximity effect in F/TSC junctions. The angle $\theta_{md}$ can be reduced with either rotation of $m$ or $d$-vector that dictates the proximity effect in the junction accordingly~\cite{Terrade2013}. It is most likely that the rotation of $m$ of SRO113 is responsible for first strong suppression in $V_3$ up to coercive field of SRO113-layer $\approx$ 200 mT~\cite{Anwar2016}. It may also be the reason of hysteresis in $V_3$ below 200~mT; above 200~mT, $m$ is saturated and fully aligned along the $H_{\rm in}$. In the results $\theta_{md}$ decreases up to coercive field and suppresses the proximity effect. On the other hand, at the same time, the applied field $H_{\rm in}$ tends to align the $d$-vector along the $c$-axis. Therefore, $\theta_{md}$ never becomes zero and proximity effect is weakly suppressing with increasing $H_{\rm in}$ above 200~mT. In this configuration of $d$-vector at the interface as illustrated in Fig.~\ref{IV_RT}(c), the $H_{\rm out}$ cannot rotate the $d$-vector, as spins of Cooper pairs are aligned out-of-plane. Therefore, the $V_3$ suppresses monotonically with $H_{\rm out}$. Of course, temperature change between 2~K and 0.3~K cannot change $m$ or $d$-vector, thus, there is no anomalous suppression in $V_3$ as a function of temperature.

The rotation of the $d$-vector under the influence of an external magnetic field is well known for superfluid ${}^3$He-A phase~\cite{Lee1997} analogous to chiral $p$-wave spin-triplet superconductivity anticipated in SRO214. In case of the standard chiral-$p$-wave scenario for SRO214, the $d$-vector should be fixed along the $c$-axis because of spin-orbit coupling. Possibility of rotation of the $d$-vector to an in-plane direction under $c$-axis magnetic field is discussed in the NMR study for $H~\parallel~c$~\cite{Murakawa2004}. However, at SRO113/SRO214 interface, broken inversion symmetry and out-of-plane exchange field of SRO113 can significantly affect the spin-orbit coupling. That may result into easier rotation of the $d$-vector along the in-plane direction. This scenario explains our results well. Theoretically, Annett {\it et al.},~\cite{Annett2008} predicted that the $d$-vector rotation by magnetic field along the $c$-axis is accompanied by chiral to nonchiral phase transition depending on spin dependent effective pairing interaction. However, for a proximity effect in a junction based on a spin-triplet superconductor, the crucial parameter is the orientation of the $d$-vector~\cite{Terrade2013}, whereas, the transition in the orbital part may not considerably modulate the proximity effect unless the junction size is of the order of the typical size of a chiral domain.

Effect from the orbital part: SRO214 exhibits multi-component ($p_x+ip_y$) superconducting order parameter~\cite{Anwar2017}. That may also explain this anisotropic behaviour of $V_3$. The component $p_x$ can be significantly suppressed compared to the $p_y$ when magnetic field $H_{\rm in}$ is parallel to $y$ and vice versa~\cite{Agterberg1998,Kaur2005,Ishihara2013}. However, the fields along $z$-direction, perpendicular to the both $p_x$ and $p_y$ components, can simultaneously affect both components of the order parameter. In this way, the differential conductance should depend differently on the in-plane and out-of-plane applied magnetic fields. Such anisotropic effect corresponding to the order parameter of the bulk SRO214 should emerge mainly in the SRO214 side and that should result in anisotropic behaviour either only in $V_2$ or both $V_2$ and $V_3$. Thus, our observed anomalous anisotropic behaviour only in $V_3$ cannot be explained by the above mentioned non-chiral scenario. Rather, this observation supports that the spin-triplet proximity effect is controlled with the change in $\theta_{md}$ due to the relative rotation of m and d-vector. More experimental and theoretical works are needed to understand and manipulate the rotation of $\theta_{md}$.

\section{Conclusion}

We studied long-range proximity effect in Au/SRO113/SRO214 double barrier junctions. Two distinct Andreev reflection features are observed in differential conduction vs bias voltage coming from two SRO113/SRO214 and Au/SRO113 interfaces with the induction of spin-triplet Cooper pairs in 15-nm thick SRO113 from SRO214. The transition of Au/SRO113 interface is suppressed anisotropically with externally applied magnetic fields. This anomalous anisotropic behaviour cannot be explained with vortex dynamics since differential conductance as a function of temperature and out-of-plane applied fields if qualitatively similar. However, conductance of SRO113/SRO214 interface suppressed faster when field is applied along the in-plane direction. Our results will stimulate the theoretical work to understand the correlation between unconventional superconductivity and ferromagnetism, $p$-wave proximity effect and applications to initiate the Superconducting Spintronics.   

\section*{Methods}

Single crystals of SRO214 were grown in Kyoto using a floating-zone method~\cite{Mao2000}. Some parts of the SRO214 crystals tend to contain Sr$_3$Ru$_2$O$_7$, SRO113, as well as Ru-metal inclusions. These inclusions are unavoidable in order to eliminate Ru deficiencies and obtain $T_{\rm c}$ close to the intrinsic $T_{\rm co}$ $\approx$ 1.5~K. To fabricate our superconducting junctions, we carefully choose the parts of SRO214 crystals that do not contain impurities but with a slightly lower $T_{\rm c}$ $\approx$ 1.23~K. Ferromagnetic SRO113 thin films with thickness of 15-nm were epitaxially deposited using pulsed laser deposition on the cleaved $ab$-surface of the SRO214 substrates (3 $\times$ 3 $\times$ 0.5 mm$^3$) in Seoul. The details of SRO113 thin film deposition can be found elsewhere~\cite{Anwar2015,Anwar2016}. Immediately after the growth of SRO113 film, a 5-nm thick Ti adhesive layer and a 20-nm thick Au capping layer were deposited {\it ex-situ} by dc-sputtering. 

Double interface Au/SRO113/SRO214 junctions shown schematically in Fig.~\ref{device_RT}a were fabricated in RIKEN. First, 25~$\times$~25~$\mu$m$^2$ and 10~$\times$~10~$\mu$m$^2$ pads of SRO113 were prepared on flat surfaces using photolithography and Ar ion etching. Then, an insulating 300-nm thick SiO$_x$ layer was sputtered. Finally, a 600-nm thick Au top electrode was deposited by electron beam evaporation with junction areas of 20~$\times$~20~$\mu$m$^2$ and 5~$\times$~5~$\mu$m$^2$ over 25~$\times$~25 $\mu$m$^2$ and 10~$\times$~10~$\mu$m$^2$ SRO113 pads, respectively. Note, a 30-nm thick neck part of SRO214 substrate was prepared with over-etching during SRO113 layer etching to define a bottom superconducting electrode, as shown in Fig.~\ref{device_RT}(a). Detailed descriptions of these processes are given in Ref.~\cite{Anwar2016}.

Electrical transport measurements were performed using the four-point technique with two contacts ($I_{+}$, $V_{+}$) on the Au top electrode and the other two contacts ($I_{-}$, $V_{-}$) connected directly to the side of the SRO214 crystal. Transport properties were studied down to 300 mK using a ${}^3$He cryostat equipped with a vector superconducting magnet.

\section*{Acknowledgement}
We are thanful to D. Manske, M. Couco and S. Arif for fruitful discussions. This work is supported by the JSPS KAKENHI projects Topological Quantum Phenomena (JP22103002 and JP25103721) and Topological Materials Science (JP15H05851, JP15K21717 and JP15H05852), JSPS KAKENHI 17H04848, as well as by JSPS-EPSRC core-to-core program ”Oxide-Superspin (OSS)” . MSA is supported as an International Research Fellow of the JSPS.

\section*{Author contributions statement}
M.S.A., S.Y. and Y.M. designed the experiments. S.R.L. grown the SRO113 thin films and preformed XDR experiments under supervision of T.W.N.. R.S. and M.S.A. fabricated the devices. C. W. grew the single crystals. M.S.A and M.K. prepared the SRO214 substrates and performed the measurements under supervision of Y.M.. M.S.A, M.K. S.Y. and Y.M. wrote the manuscript. All authors contributed in data analysis, discussion and reviewing the manuscript.

\section*{Additional information}
The authors declare no competing interests.

\end{document}